\documentstyle[twoside,fleqn,espcrc2]{article}
\begin{document}

\def\NPB#1#2#3{{\sl Nucl. Phys.} \underbar{B#1} (#2) #3}
\def\PLB#1#2#3{{\sl Phys. Lett.} \underbar{#1B} (#2) #3}
\def\PRL#1#2#3{{\sl Phys. Rev. Lett.} \underbar{#1} (#2) #3}
\def\PRD#1#2#3{{\sl Phys. Rev.} \underbar{D#1} (#2) #3}
\def\CMP#1#2#3{{\sl Comm. Math. Phys.} \underbar{#1} (#2) #3}
\def\hepth#1{{hep-th/#1}}
\def\hepph#1{{hep-ph/#1}}
\def\del{\partial}
\def\bar{\overline}
\def\bC{{\bf C}}
\def\bZ{{\bf Z}}
\def\ie{{\it i.e.}}
\def\eg{{\it e.g.}}
\def\til{\widetilde}
\def\tA{{\til A}}
\def\tF{{\til F}}
\def\DD{{\cal D}}
\def\MM{{\cal M}}
\def\OO{{\cal O}}
\def\WW{{\cal W}}
\def\tQ{{\til Q}}
\def\tq{{\til q}}
\def\vev#1{{\langle #1\rangle}}
\def\bea{\begin{eqnarray}}
\def\eea{\end{eqnarray}}
\def\nn{\nonumber}
\def\be{\begin{equation}}
\def\ee{\end{equation}}

\title{Dualities in Supersymmetric Field Theories}

\author{Philip C. Argyres\address{Newman Laboratory,
        Cornell University, Ithaca NY 14853, USA}}
       
\begin{abstract}
These lectures\thanks{presented at the 33rd Karpacz Winter School
``Duality: Strings and Fields'' (Poland, February 1997).} briefly
introduce dualities in four-dimensional quantum field theory, and
summarize results found in supersymmetric field theories.  The first
lecture describes physical aspects of electric-magnetic (EM),
strong-weak coupling (S), and infrared (IR) dualities.  The second
lecture focuses on results and conjectures concerning S-duality in
$N{=}2$ supersymmetric gauge theories.  The third lecture discusses
IR-dualities and their relation to S-duality in $N{=}1$ supersymmetric
field theories.
\end{abstract}

\maketitle

\section{WHAT IS DUALITY?}

The word ``duality'' has been used to describe many things in physics.
We will examine those dualities which broadly deal with the quantum
equivalence of gauge field theories.

\subsection{EM-duality}

The simplest example of such an equivalence is EM duality, which is
apparent classically.

Maxwell's equations with electric and magnetic sources
	\be
	d*F = *j_e, \qquad dF  = *j_m,
	\ee
are invariant under the transformation
	\be\label{EMduality}
	F \leftrightarrow *F, \qquad
	j_e^\nu \leftrightarrow j_m^\nu ,
	\ee
where $*F_{\mu\nu} \equiv {1\over2}\epsilon_{\mu\nu\rho\sigma}
F^{\rho\sigma}$.

This duality of the classical equations of motion can be shown to hold
quantumly as well.  To see this, consider a $U(1)$ theory with
coupling
	\be
	\tau = {\vartheta\over2\pi} + i {4\pi\over g^2}
	\ee
(the theta-angle is important in a $U(1)$ theory when there are both
electric and magnetic sources).  The action can be written
	\be
	S = \int d^4x\, {\tau\over 32\pi i} (F + i{*F})^2 + h.c.
	\ee
We compute physical quantities in this theory as a path integral over
all gauge potential configurations $\int\DD A_\mu e^{iS}$.  This can
be rewritten as a path integral over field strength configurations as
long as we insert the Bianchi identity as a constraint: $\int\DD F 
\DD\tA e^{iS'}$, where
	\be
	S' = S + {1\over8\pi} \int d^4x\, \tA\wedge dF .
	\ee
Here $\tA_\mu$ is a Lagrange multiplier enforcing the Bianchi
identity, and whose normalization is chosen so that it couples to a
monopole with charge one.  Performing the Gaussian functional integral
over $F$, we find an equivalent action, $\til S$, for $\tA$:
	\be
	\til S = \int d^4x\, {1\over32\pi i} \left( -1\over\tau\right)
	(\tF + i{*\tF})^2 ,
	\ee
where $\tF = d\tA$ is the field strength of $\tA$.  These steps can be
performed just as easily in supersymmetric Maxwell theory.

Thus the free $U(1)$ gauge theory with coupling
$\tau$ is quantumly equivalent to another such theory with coupling
$-1/\tau$.  This is the EM-duality ``symmetry''.  It is not really a
symmetry since it acts on the couplings---it is an equivalence between
two descriptions of the physics.  The EM-duality transformation ($S$)
and the invariance of the physics under $2\pi$ shifts of the
$\vartheta$-angle ($T$),
	\be
	S:\ \tau \rightarrow -1/\tau, \qquad
	T:\ \tau \rightarrow \tau{+}1 ,
	\ee
generate the group $SL(2,\bZ)$ of duality transformations:
	\be\label{sltz}
	\tau \rightarrow {a\tau + b \over c\tau + d}, 
	\qquad \pmatrix{a&b\cr c&d\cr} \in SL(2,\bZ). 
	\ee

Because the dual $U(1)$ potential couples to monopoles with unit
strength, we see that the EM-duality transformation interchanges what
we mean by electric and magnetic charges.  Under this transformation,
a massive dyonic source with electric and magnetic charges $(n_e,n_m)$
in the original description couples to the dual $U(1)$ with charges
$(n_m, -n_e)$.  The minus sign arises because the square of a duality
transformation (\ref{EMduality}) is charge conjugation.  The effect of
the $T$ transformation on the charges is $(n_e,n_m) \rightarrow
(n_e{+}n_m, n_m)$, as follows from the Witten effect \cite{WittenEffect}.

This description of EM-duality might make it seem trivial, since all
it deals with is the equivalence of free field theories coupled to
classical (massive) sources.  Its importance stems from the fact that
some interacting field theories have vacua with massless photons and
no other massless charged matter.  The coupling of the effective
low-energy Maxwell theory is determined by the mass of the lightest
charged particle.  EM-duality is then simply a redundancy in the
description of the IR-effective action of this theory which is
important to take into account in determining it, as we will see in
$N{=}2$ supersymmetric examples in the next lecture.

\subsection{S-duality}

S-duality is the exact quantum equivalence of two apparently
(classically) different field theories.  This equivalence often takes
the form of the identification of a theory with specific deformations
of that theory by exactly marginal operators.  These deformations can
be viewed as transformations on the classical space of couplings of
the theory, and the duality group is the group of such
transformations.  Since the elements of the duality group are supposed
to connect equivalent theories, the quantum coupling space is the
classical one divided by the action of the duality group.

A famous example \cite{OliveMontonen,WittenOlive,Osborn} is the
strong-weak coupling duality of $N{=}4$ supersymmetric Yang-Mills
under which theories with gauge couplings $g^2$ and $1/g^2$ are
identified.  There is by now rather compelling evidence for this
duality in field theory having to do with the spectrum of BPS
saturated states as described in J. Gauntlett's lectures, and
with the low-energy effective action \cite{VafaWitten,SeibergWittenII}.

In this case the exactly marginal operator corresponds to changing the
coupling, since its exact beta-function vanishes.  Classically, the
coupling can take any positive value, and so the coupling constant
space has the topology of an open interval: $(0,\infty)$.  Under the
S-duality identification, however, we see that the topology is
modified quantumly to that of a half-closed interval $(0,1]$.

Another quantum identification of the coupling space of the $N{=}4$
theory is the angularity of the theta angle: $\vartheta \simeq
\vartheta + 2\pi$.  The classical parameter space is then the upper-half 
plane $\MM_{cl} = \{{\rm Im}\tau > 0 \}$, where $\tau \equiv
(\vartheta/2\pi) + i (4\pi^2/g^2)$.  Combining the theta-angle
identification visible at weak coupling with $g^2 \simeq 1/g^2$ gives
the quantum coupling space $\MM = \MM_{cl}/SL(2,\bZ)$, where
$SL(2,\bZ)$ acts on the coupling as in (\ref{sltz}).  The topology of
$\MM$---the fundamental domain of $SL(2,\bZ)$ with
identifications---is that of an open disk.  The duality group,
$SL(2,\bZ)$, can be presented as the group with two generators $S$ and
$T$ satisfying the relations $S^2=(ST)^3=1$.  This group clearly
encodes more information than just the topology of $\MM$.  In
particular, the generators $S$ and $ST$ do not act freely on
$\MM_{cl}$, but have fixed points which are $\bZ_2$ and $\bZ_3$
orbifold points of $\MM$, if we assign $\MM_{cl}$ a flat metric.  The
relations satisfied by these generators then describe the holonomy in
$\MM$ around these orbifold points.

Since we are free to make coordinate changes on $\MM_{cl}$, though the
topology of $\MM$ certainly has an invariant meaning, it may not be
clear that the holonomies around points in $\MM$ are physically
meaningful.  In particular, to define holonomies, one needs a
connection.

A natural connection on the coupling space $\MM$ does exist
\cite{JackOsborn,Sonoda,Dolan}.  The idea is simply that $n$-point
correlators of the operators $\OO^i$ conjugate to the couplings
$\tau_i$ which coordinatize $\MM$ transform as rank-$n$ tensors under
coordinate changes on $\MM$.  The derivative of an $n$-point
correlator with respect to the coupling $\tau_i$ can be computed in
terms of the $(n{+}1)$-point correlator with an insertion of $\OO^i$.
This insertion must be regularized when it coincides with the
space-time position of another operator in the correlator, giving, due
to Schwinger terms in the operator product expansion (OPE)
\be\label{ope} 	\OO^i(x) \OO^j(0) = c^{ij}_k \delta(x) \OO^k(0) +
\ldots , 	\ee a sum of $n$-point correlators with $\OO^k$
insertions replacing the original operators.  Since all the other
terms transform covariantly under coordinate transformations on $\MM$,
it must be that the $c^{ij}_k$'s transform in such a way as to cancel
the non-covariant part from the coordinate derivative
$\del/\del\tau_i$, \ie\ as a connection on $\MM$.  This connection
(the $c^{ij}_k$'s) arises as finite counterterms upon subtracting the
singular parts of the OPE in a given RG scheme.

It is a complicated business to compute the above connection; moreover
the OPE does not in general close on the original conjugate operators,
necessitating enlarging $\MM$ to include all of theory space.
However, in the case at hand, namely four-dimensional supersymmetric
field theories with exactly marginal couplings in the superpotential,
the situation dramatically simplifies.  Because the conjugate
operators are exactly marginal, they remain in the theory at the IR
fixed point; because they are chiral operators, their OPEs (at the IR
fixed point) have no singular parts due to the proportionality of
their scaling dimensions to their $R$-charges \cite{Seiberg9411}.
Thus the connection vanishes, and $\MM$ is locally flat.  

The duality group elements which do not act freely on $\MM_{cl}$
therefore do indeed encode a physical holonomy of operators parallelly
transported in $\MM$ around their fixed points.  The known
(conjectured) S-duality groups for $N{=}2$ supersymmetric field
theories will be discussed in the next lecture.

\subsection{IR-duality}

IR-duality refers to the equivalence of two theories in the infrared,
\ie\ at long distances or low energies.  In the context of supersymmetric
gauge theories, this is sometimes refered to as ``non-Abelian
duality'' or ``$N{=}1$ duality''.  However, more generally, IR-dual
theories are simply theories that belong to the same RG universality
class.

The reason we give a special name to theories in the same universality
class is that in the case of supersymmetric gauge theories, the
resulting IR-dual pairs have properties reminiscent of the $\tau 
\leftrightarrow -1/\tau$ transformation in EM and S-dualities.
In particular, it is found that if both theories in the IR-dual pair
are asymptotically free (AF), then their strong-coupling scales
$\Lambda_1$ and $\Lambda_2$ will vary inversely:
	\be
	\Lambda_1^{b_1} \Lambda_2^{b_2} \sim \mu^{b_1+b_2},
	\ee
where $b_1$ and $b_2$ are specified constants, and $\mu$ is
an arbtrary matching scale.  

More strikingly, there are IR-dual pairs one of which is AF and the
other IR-free, which have different gauge groups.  Restated, this
means that certain AF theories which are well-described at short
distances in terms of a gauge group $G_1$ give rise to long
wave-length physics which is well-described by a different gauge group
$G_2$.  Furthermore, there is no clear sense in which the gauge bosons
of $G_2$ are related to those of $G_1$; in particular, $G_2$ is not
simply a subgroup of $G_1$.

We will discuss the evidence for IR-duality and its possible
connection to S-duality in the third lecture.

\subsection{Extensions}

So far we have discussed EM-duality in free $U(1)$ gauge theories,
S-duality in theories with exactly marginal couplings, and IR-duality
of theories with couplings which undergo RG flow.  The concepts as
introduced have been quite general, and so might be expected to apply
more broadly.

In particular, none of these dualities relies in principle on
supersymmetry.  We formulated EM-duality in Maxwell theory.  The
description of S-duality as the quantum modification of the topology
(and holonomies) of the parameter space of couplings and IR-dualities
as universality classes of gauge theories show them to be potentially
generic phenomena in field theories.  However, at present only
supersymmetry gives us the control over the strong coupling dynamics
of gauge theories necessary to give evidence for S- and IR-dualities.

Also, nothing in the description of these dualities limits them to
four dimensions.  Indeed, we have seen extensions of EM-duality to
other dimensions in F. Quevedo's lectures.  Likewise, analogs of
S-duality in supergravity and string theories in various dimensions
are the subject of many other lectures at this school.

\section{N=2 S-DUALITY}

This lecture reviews known S-dualities in $N{=}2$ supersymmetric gauge
theories.  We start by setting notation with a quick review of $N{=}2$
superQCD (see the talks of A. Klemm for more details).

There are two $N{=}2$ supermultiplets which concern us, the vector
multiplet and the hypermultiplet.  An $N{=}2$ vector multiplet is made
up of an $N{=}1$ vector multiplet $V$ and an $N{=}1$ chiral multiplet
$\Phi$ in the adjoint.  We refer to the complex adjoint scalar
component of $\Phi$ as the Higgs field.  An $N{=}2$ vector multiplet
is made up of two $N{=}1$ chiral multiplets $Q$ and $\tQ$ transforming
in a representation $R$ and $\bar R$ respectively, and we refer to
their scalar components as quarks.

The only relevant or marginal couplings allowed by $N{=}2$
supersymmetry are hypermultiplet masses and the gauge coupling
$\tau$.  The one-loop beta function for $\tau$ is
	\be
	\beta_{1\ell} = - T(Ad) + \sum_i T(R_i)
	\ee
where $T(R)$ is the index of the $R$ representation of the gauge group
($Ad$ is the adjoint representation), and the sum is over
hypermultiplets in representations $R_i$.  Thus, for example, for
$SU(n_c)$ with $n_f$ hypermultiplets in the fundamental
representation, $\beta_{1\ell} = -2n_c + n_f$.

To look for S-duality, we want an exactly marginal operator.  In the
above example, we can arrange for the beta function to vanish by
taking $n_f = 2n_c$.  However, this is only the one-loop beta
function; how can we show that $\beta=0$ exaxctly?

\subsection{Exactly marginal operators}

A non-renormalization theorem \cite{Seiberg88,Seiberg93} states that
the beta function is given exactly by its one-loop value if
$\beta_{1\ell} \le 0$.  (In IR-free theories where $\beta_{1\ell} >
0$, non-perturbative contributions to the beta function are allowed,
and play a crucial role in the exact solution for the low-energy
effective action of $N{=}2$ superQCD \cite{SeibergWittenI}.)  However,
since beta functions are RG scheme-dependent, one must ask for which
scheme is this non-renormalization theorem valid?

The supersymmetric non-renormalization theorems are always valid in a
scheme where the ($N{=}1$) superpotential is a holomorphic function of
the couplings.  This is the case for Wilsonian effective actions in
which there is no wave-function renormalization---that is, the kinetic
terms are not rescaled to their canonical normalization after an RG
transformation.  Physical beta functions (reflecting the
scale-dependence of the physical couplings) are measured in a scheme
with wave-function renormalization, though.  Translating to such a
scheme effectively adds higher-loop and non-perturbative corrections
to the one-loop beta-function through the wave function
renormalization of fields propagating in the loop.  The exact
wave-function renormalization is encoded in the anomalous dimensions
$\gamma_i$ of the hypermultiplets.

Reasoning along these lines was used in \cite{NSVZ} to show that the
exact physical beta function is proportional to
	\be\label{exbeta}
	\beta_\tau \propto \beta_{1\ell} + \sum_i \gamma_i T(R_i).
	\ee
Similarly, for an operator
	\be\label{exlam}
	\delta\WW = \lambda \prod_i (\Phi_i)^{r_i}
	\ee
in the superpotential (in $N{=}1$ language), the exact beta-function 
for its coupling $\lambda$ is
	\be
	\beta_\lambda \propto 3 - \sum_i r_i(1+\gamma_i).
	\ee

Now, following \cite{LeighStrassler}, we look for fixed point theories
by solving for points in $\tau$-$\lambda$ coupling space where
$\beta_\tau=\beta_\lambda=0$.  These are two conditions on two unknown
parameters, and so generically have isolated fixed-point solutions.
However, if we can arrange that $\beta_\tau \propto \beta_\lambda$,
then there is only one independent condition on two unknowns, and so
there will typically be a line of fixed points and thus an exactly
marginal operator (which moves us along this line).

To apply this strategy for finding exactly marginal operators to
$N{=}2$ theories, consider the $N{=}1$ superQCD theory with an adjoint
chiral superfield $\Phi$ and pairs of chiral ``quark'' superfields
$Q^i$ and $\tQ_i$ in representations $R_i$ and $\bar R_i$ (\ie\ the
field content of an $N{=}2$ superQCD theory).  In addition to the
gauge coupling $\tau$, we add superpotential terms $\WW =
\sum_i\lambda_i Q^i \Phi \tQ_i$.  By the global symmetries of this
theory, the anomalous dimensions of the quarks and anti-quarks are the
same: $\gamma_{Q^i} = \gamma_{\tQ_i} \equiv \gamma_i$.  {}From
(\ref{exbeta}) and (\ref{exlam}) we then get
	\bea
	\beta_\tau &\propto& (\gamma_\Phi{-}1) T(Ad) 
	+ \sum_i (1{+}2\gamma_i) T(R_i), \nn\\
	\beta_{\lambda_i} &\propto& \gamma_\Phi + 2 \gamma_i,
	\eea
implying the existence of a fixed line and an exactly marginal
operator precisely when the one-loop beta function vanishes, $\sum_i
T(R_i) = T(Ad)$, since then $\beta_\tau \propto \sum_i T(R_i)
\beta_{\lambda_i}$.  In addition, one can check that the fixed line is
approximately $\tau=\lambda_i$ at weak coupling which are the $N{=}2$
supersymmetric values of the couplings.

\subsection{Low-energy effective actions and S-duality groups}

Evidence for S-duality in these scale-invariant $N{=}2$ theories can
be found by looking at the exact low-energy effective action on the
Coulomb branch of their moduli spaces.

Classically, the Coulomb branch $\MM$ is the manifold of degenerate
vacua where the quarks have zero vevs while the Higgs field takes
values in the (complexified) Cartan subalgebra of the gauge group.
This breaks the gauge group as $G \rightarrow U(1)^r$ where
$r=$rank$(G)$.  Thus (locally) $\MM \simeq \bC^r$, and we will denote
its coordinates by $\vev{\phi}_k$, $k=1, \ldots, r$.

Quantumly, a non-renormalization theorem implies these flat directions
are not lifted.  Note that the point $\vev{\phi}_k = 0$ will be the
scale-invariant (fixed-point) vacuum, but that moving away from it
along the Coulomb branch to points with $\vev{\phi}_k \neq 0$
spontaneously breaks the scale invariance of the underlying theory and
generically flows to the IR-free $U(1)^r$ Maxwell theory.  The
couplings $\tau_{jk}$ of this low-energy effective $U(1)^r$ theory can
receive quantum corrections.

The $U(1)$ $SL(2,\bZ)$ EM-duality group generalizes straight-forwardly
to an $Sp(2r,\bZ)$ EM-duality group in the $U(1)^r$ theory.  (Note
that the appearance of this low-energy EM-duality has nothing to do
with the S-duality we are looking for!)  Seiberg and Witten
\cite{SeibergWittenI} observed that this EM-duality implies that the
effective couplings $\tau_{jk}$ need not be a single-valued function
on $\MM$ (the Coulomb branch), but can be a section of an $Sp(2r,\bZ)$
bundle on $\MM$.  We encode the $Sp(2r,\bZ)$ ``periodicity'' of
$\tau_{jk}$ geometrically by taking $\tau_{jk}$ to be the complex
structure of a $T^{2r}$ torus which varies over $\MM$.  It is shown in
\cite{DonagiWitten} that the total space of $T^{2r}$ fibered over
$\MM$ is the phase space of an integrable system.  Furthermore, in
every known example it turns out that this auxiliary torus $T^{2r}
{=}$Jac$(\Sigma_r)$, the Jacobian torus of a genus-$r$ Reimann surface
$\Sigma_r$, though no explanation of this fact is known.

The solutions for the low-energy effective action on the Coulomb
branch are usually presented in terms of the $\Sigma_r$.  No
systematic procedure is known for generating these families of Riemann
surfaces.  The solutions have so far been found either by making
simple guesses for the form of the surfaces or for the integrable
system, or from type IIB string compactifications as reviewed in A.
Klemm's lectures in this school or, most recently, from brane
configurations in type IIA string theory or M-theory
\cite{Witten9703}.

The scale-invariant $N{=}2$ theories whose associated Riemann surfaces
were known (before \cite{Witten9703}) are the theories with a single
$SU$, $SO$, or $Sp$ gauge group and quarks in the fundamental
(defining) representation \cite{APS,AS}.  It is found in all these
cases that these curves depend on the marginal coupling $\tau$ in such
a way that they are invariant under identifications of $\tau$ under a
discrete group isomorphic to $\Gamma^0(2) \subset SL(2,\bZ)$.  This is
evidence that the S-duality group of all these theories is isomorphic
to $\Gamma^0(2)$.

$\Gamma_0(2)$ is the subset of $SL(2,\bZ)$ matrices with even upper
off-diagonal entry, or equivalently, is isomorphic to the subgroup of
$SL(2,\bZ)$ generated by $S$ and $T^2$.  The only relation satisfied
by these generators is $S^2=1$, thus characterizing $\Gamma^0(2)$ more
abstractly as the group freely generated by $S$ and $T'$ with the one
relation $S^2=1$.  Since the S-duality group is the group we divide
the classical coupling space by, we expect a single point with a
$\bZ_2$ holonomy in the quantum coupling space at the fixed point of
the $S$-transformation's action.  Indeed, the fundamental domain of
$\Gamma^0(2)$ in the $\tau$ upper half-plane can be taken to be the
region $-1{\le}$Re$\tau{\le}1$ and $|\tau| \ge 1$.  The boundaries of
this domain are identified so that there is a $\bZ_2$ orbifold point
at $\tau = i$ and ``cusps'' with no holonomy at $\tau=i\infty$ (weak
coupling) and $\tau = \pm1$ (ultra-strong coupling).

An important feature of this duality group is that it does not remove
all the ultra-strong coupling points from the quantum coupling space.
An important exception is $SU(2)$ $N{=}2$ superQCD with four
fundmental flavors.  For this theory it was shown
\cite{SeibergWittenII} that the low-energy effective theory at the
ultra-strong coupling point is in fact equivalent to the weak coupling
limit of the same effective action.  This is evidence that the
S-duality group in this case is enlarged to the full $SL(2,\bZ)$
group.  In the other cases (with higher-rank gauge groups) no weakly
coupled description of the ultra-strong couplings points is known.  In
particular they are not the weak coupling limit of the same theories,
so the S-duality groups are not ``secretly'' enlarged to $SL(2,\bZ)$,
if they are enlarged at all.

\subsection{Recent developments}

Since these lectures were given, the list of conjectured $N{=}2$
S-duality groups (based on equivalences of exact low-energy effective
actions on Coulomb branches) has increased dramatically due to the
work of Witten \cite{Witten9703}.  He argues that 5-brane
configurations in M-theory in which two of the dimensions of the
5-brane world-volume form a Riemann surface $\Sigma$ and the other 3+1
are flat give rise to two natural families of $N{=}2$ four-dimensional
superQCD theories.  {}From its embedding in the 11-dimensional
space-time of M-theory, $\Sigma$ inherits a metric structure, and, in
particular, has metric infinities.  For the scale-invariant theories,
the values of the couplings can be read off from the relative
distances of these metric infinities in two transverse dimensions, say
$x^6$ and $x^{10}$, in the 11-dimensional space-time.

The ``cylindrical'' family of $N{=}2$ models constructed in this way
has gauge group $G = \prod_{i=1}^n SU(k_i)$, $n{-}1$ hypermultiplets
in the $({\bf k_i, k_{i+1}})$ bi-fundamental representations, as well
as $f_i$ hypermultiplets in the fundamental of each $SU(k_i)$
factor.  This is a scale-invariant model with $n$ exactly marginal
couplings $\tau_i$ if
	\be
	2k_i - k_{i-1} - k_{i+1} = f_i, \qquad
	k_0 \equiv k_{n+1} \equiv 0 .
	\ee
In these models the $x^6$ coordinate is infinite and the $x^{10}$
coordinate lives on a circle making a cylindrical transverse space.
$\Sigma$ has $n{+}1$ metric infinities, and their relative positions
on the $x^6$-$x^{10}$ cylinder give the $n$ complex coupling
parameters of the model.

This gives a direct description of the quantum space of couplings as
the moduli space of complex structures of a cylinder with $n{+}1$
indistinguishable punctures.  The indistinguishablity of the punctures
can be seen in an equivalent type IIA string picture by exchanging the
positions of any two infinite five-branes.  The S-duality group can be
extracted from this description of the moduli space, and can easily be
seen to be isomorphic to $\Gamma^0(2)$ when $n{=}1$.  The more general
case has an S-duality group generated by translations $T_i$ of each
puncture around the cylinder, as well as exchanges $S_{i,i+1}$ of
pairs of punctures.  Dividing the classical coupling space by this
group gives rise to a complicated set of orbifold submanifolds and
associated holonomies.  Also, two punctures colliding corresponds to a
point of ultra-strong coupling.  In particular, when applied to the
$SU(2)$ scale-invariant theory, one only sees in this manner the
$\Gamma^0(2)$ S-duality, and not the larger $SL(2,\bZ)$ group.

The second, ``elliptic'', family of models arises upon
compactifying the $x^6$ direction on a circle.  The scale-invariant
$N{=}2$ theories have gauge group $G= SU(k)^n$ with $n$
hypermultiplets in bi-fundamental representations.  The Riemann
surface has $n$ metric infinities puncturing the $x^6$-$x^{10}$ torus,
so the quantum space of couplings is the moduli space of complex
structures of a torus with $n$ indistinguishable puntures.  When
$n{=}1$ these are the $SU(k)$ $N{=}4$ theories with S-duality group
$SL(2,\bZ)$.  With $n>1$, however, there are still ultra-strong
coupling points in these models which have no known weak-coupling
description.

\section{IR-DUALITY AND N=1 S-DUALITY}

This lecture briefly reviews the IR-dualties found in $N{=}1$
supersymmetric gauge theories and explores the connection between
IR-duality and S-duality.

Recall the first example \cite{Seiberg9411} of a series of IR-dual
$N{=}1$ theories, which gives the flavor for most of the IR-dual
theories found to date.  These IR-dual pairs consist of an
``electric'' theory which is $SU(n_c)$ superQCD with $n_f \ge n_c+2$
fundamental flavors $Q^i$, $\til Q_i$ and no superpotential, and a
``magnetic'' $SU(\til n_c)$ superQCD with $n_f$ fundamental flavors
$q_i$, $\til q^i$, a gauge singlet field $M^i_j$, and a superpotential
$\WW_{mag} = (1/\mu) Mq\til q$.  The two theories have the same IR
behavior if
	\be
	\til n_c = n_f - n_c .
	\ee
Identifying the magnetic singlet $M$ with a composite meson chiral 
superfield in the electric theory, $\mu$ in the magnetic superpotential
becomes a matching scale between the electric and magnetic theories.
If $\Lambda_e$ and $\Lambda_m$ are the strong-coupling scales of
the electric and magnetic theories, respectively, then they are
related by
	\be
	\Lambda_e^{3n_c-n_f} \Lambda_m^{3\til n_c - n_f} \sim \mu^{n_f}.
	\ee
Note that the one-loop beta fuction $\beta_{1\ell} = -3n_c +n_f$.
An important feature of this duality is that for $n_f/3 \le n_c, 
\til n_c \le 2n_f/3$ both the electric and magnetic theories are AF,
and flow to a non-trivial fixed point in the IR.  There is a ``self-dual''
theory at $n_c = \til n_c = n_f/2$ where the two gauge groups are the
same (though the theories differ by singlet degrees of freedom).  Moreover,
starting from the self-dual theories, one can flow down to all the
other dual pairs by turning on appropriate relevant operators in the
superpotential.

Most of these basic features are shared by the many known IR-dual
pairs; for a fairly complete (as of early 1997) listing of the known
IR-dualities and their properties, see \cite{Brodie}.  One striking
uniformity is that all known IR-dual series, except for one
\cite{Pouliot}, have self-dual pairs from which all the other dual
pairs in a series can be derived by deformation by relevant operators
in the superpotential.

The evidence for the IR-equivalence of these theories is based on the
equivalence of their chiral ring of operators (\ie\ the composite
massless chiral superfields, more or less) and the equivalence of the
theories under a large set of relevant deformations.  The matching of
the chiral rings means, in practice, that anomalous dimensions of
operators at the IR fixed points match, as do the moduli spaces of
vacua of the two theories.

Unlike the case of S-duality, which by its very nature involves strong
couplings, one can imagine a proof of IR-duality using existing field
theory techniques.  Recalling that IR-dualities are just universality
classes of gauge theories, one needs only show that two theories are
connected by an irrelevant deformation to show they are IR-dual.  The
standard strategy to do this is to find an UV fixed-point theory which
flows to both members of the dual pair upon deformation by certain
relevant operators.  Then one needs to show that the difference
between those relevant operators is itself irrelevant.  One must be
careful, however, since an operator which looks irrelevant near the UV
fixed point, may become relevant as it flows to the IR (so-called
``dangerously irrelevant operators'').  In the supersymmetric case one
has some control over this subtlety.  At least for non-chiral
theories, by a Witten index argument one can often regulate the theory
in the IR by putting it in finite volume (so it becomes supersymmetric
quantum mechanics) and still keep supersymmetry unbroken so that the
vacuum energy is always zero.  As long as the operator in question
does not change the potential significantly at large field values, the
vacuum energy will depend analytically on its coupling, which together
with unbroken supersymmetry disallows any crossover in the vacuum
state at the coupling is varied.  In this way one could show that the
operator is indeed irrelevant (or, at worst, exactly marginal) in the
IR.

This strategy was pursued in the case of the $SU$ \cite{APSe}
and $SO$ and $Sp$ \cite{APSII} IR-dual superQCD series.  The UV
fixed-point theory was taken to be the related AF or scale-invariant 
$N{=}2$ superQCD.  Though it was shown how to flow down in these
theories to $N{=}1$ theories with either the electric or magnetic
gauge groups, the singlet meson fields of the magnetic theory
were not found.  It turns out \cite{AB} that for all the AF $N{=}2$
theories, the perturbation used to flow down to the $N{=}1$ dual
pairs is relevant at their IR fixed points, and so cannot be used
to prove IR-duality.  In the scale-invariant $N{=}2$ case, however,
these operators are exactly marginal in the IR \cite{LeighStrassler}.
Also, the scale-invariant $N{=}2$ theories flow to the self-dual
$N{=}1$ pairs.

In fact, all the $N{=}1$ IR self-dual pairs admit exactly marginal
deformations corresponding to turning on mass terms for the singlet
meson fields in the magnetic theory.  The fact that scale-invariant
(S-dual) $N{=}2$ superQCD theories flow to IR self-dual pairs suggests
that the IR self-dual theories might be exactly S-dual in their
marginal coupling.  Indeed, in \cite{LeighStrassler} it was argued
that upon turning on a relevant operator causing the S-dual $N{=}2$
$SU(n_c)$ theory to flow to the $n_f = 2n_c$ $N{=}1$ theory, the
exactly marginal operators of the two theories would be connected by
RG flow.  Assuming the RG flow maps the parameter spaces of these
marginal couplings in a 1-to-1 fashion, the IR (self-)duality of the
$N{=}1$ theory would follow by an action of the S-duality element
associated with the $\bZ_2$ orbifold point in the $N{=}2$ coupling
constant parameter space.

This argument depends on the assumption that the RG flow maps the
marginal coupling space of the UV theory ``into'' (in the mathematical
sense) the IR theory's marginal coupling space.  Unfortunately, it is
not at all clear why this should be the case.  In particular, the
relevant operator $\OO$ by which we flow down from the $N{=}2$ theory
to the $N{=}1$ theory will suffer some monodromy $\OO \rightarrow
\OO'$ as we transport it around an orbifold point in the $N{=}2$
coupling constant space.  The question becomes whether the operator
difference $\Delta\OO = \OO-\OO'$ is itself irrelevant or not.  For if
it is irrelevant, then the theory and its monodromic image will flow
to the same theory in the IR, implying an into mapping of the the
$N{=}2$ to $N{=}1$ parameter spaces.  But if $\Delta\OO$ is relevant,
then the mapping need not be into, potentially ``unwrapping'' the
S-duality identifications in the IR.

Nevertheless, the striking prevalence of IR self-dual theories and
their associated exactly marginal operators suggests that S-duality in
these marginal couplings is the explanantion of IR-duality.  Perhaps
with the richer $N{=}2$ S-duality groups found in the models with
product gauge groups \cite{Witten9703}, stronger evidence for the
existence of $N{=}1$ S-dualities can be found in the form of an
associated ``web'' of IR-dualities in $N{=}1$ theories with product
gauge groups.

I am grateful to A. Buchel and A. Shapere for useful
discussions.  The work was supported by NSF grant PHY-9513717.

\end{document}